\documentclass[a4paper,fleqn,usenatbib,useAMS]{mnras}

\usepackage{graphicx}	
\usepackage{amsmath}	
\usepackage{amssymb}	
\usepackage{multicol}        
\usepackage{bm}		
\usepackage{pdflscape}	

\newcommand{\Msun}{\ensuremath{\mathrm{M}_\odot}}

\newcommand{\Ni}{\ensuremath{^{56}\mathrm{Ni}}}
\newcommand{\Co}{\ensuremath{^{56}\mathrm{Co}}}
\newcommand{\Fe}{\ensuremath{^{56}\mathrm{Fe}}}
\newcommand{\Mej}{\ensuremath{M_\mathrm{ej}}}
\newcommand{\Eej}{\ensuremath{E_\mathrm{ej}}}
\newcommand{\Mni}{\ensuremath{M_\mathrm{\Ni}}}

\newcommand{\kmps}{\ensuremath{\mathrm{km~s^{-1}}}}

\usepackage[T1]{fontenc}
\usepackage{ae,aecompl}

\title[Stripped-envelope ECSNe]{
Rapidly evolving faint transients from stripped-envelope electron-capture supernovae
}

\author[T. J. Moriya and J. J. Eldridge]{
Takashi J. Moriya$^{1,2}$\thanks{takashi.moriya@nao.ac.jp}
and
J. J. Eldridge$^{3}$\thanks{j.eldridge@auckland.ac.nz}
\\
$^{1}$Division of Theoretical Astronomy, National Astronomical Observatory of Japan, 2-21-1 Osawa, Mitaka, Tokyo 181-8588, Japan \\
$^{2}$Argelander Institute for Astronomy, University of Bonn, Auf dem H\"ugel 71, D-53121 Bonn, Germany \\
$^{3}$Department of Physics, University of Auckland, Private Bag 92019, Auckland, New Zealand
}

\date{}

\pubyear{2016}

\begin{document}
\label{firstpage}
\pagerange{\pageref{firstpage}--\pageref{lastpage}}
\maketitle

\begin{abstract}
We investigate the expected rates and bolometric light-curve
properties of stripped-envelope electron-capture supernovae (ECSNe)
using stellar models from the Binary Population and Spectral Synthesis
(BPASS) code. We find that 0.8 per cent $(Z=0.020)$ and 1.2 per cent
$(Z=0.004)$ of core-collapse supernovae are stripped-envelope
ECSNe. Their typical ejecta masses are estimated to be about $0.3~\Msun$
$(Z=0.020)$ and $0.6~\Msun$ $(Z=0.004)$. Assuming ECSN explosion
properties from numerical explosion simulations, an explosion energy
of $1.5\times 10^{50}~\mathrm{erg}$ and a \Ni\ mass of $2.5\times
10^{-3}~\Msun$, we find that stripped-envelope ECSNe have a typical
rise time of around 7~days ($Z=0.020$) or 13~days ($Z=0.004$)
and peak luminosity of around $10^{41}~\mathrm{erg~s^{-1}}$ ($-13.8$~mag, $Z=0.020$)
or $7\times 10^{40}~\mathrm{erg~s^{-1}}$ ($-13.4$~mag, $Z=0.004$).
Their typical ejecta velocities are around $7000~\kmps$ ($Z=0.020$) or $5000~\kmps$
($Z=0.004$). Thus, stripped-envelope ECSNe are observed as
rapidly-evolving faint transients with relatively small
velocities. SN~2008ha-like supernovae, which are the faintest kind of
SN~2002cx-like (a.k.a. Type~Iax) supernovae, may be related to
stripped-envelope ECSNe.
\end{abstract}




\section{Introduction}
Core-collapse supernovae (SNe) are mostly the result of the collapse
of an iron core. Stars must be massive enough to form such cores and
give rise to such an event. The minimum mass is thought to be between
8 to 10~\Msun\ \citep{smartt2015}. In stars below this limit nuclear
burning progresses only to forming an ONeMg core. However such a core
may also eventually collapse and produce a SN due to electron-capture
reactions \citep[e.g.][]{miyaji1980,nomoto1984,nomoto1987}. These
events are referred to as electron-capture SNe (ECSNe). Numerical
simulations of ECSNe reveal that they explode with an explosion
energy of $\sim 10^{50}~\mathrm{erg}$ and synthesize $\sim
10^{-3}~\Msun$ of radioactive \Ni\ \citep[e.g.][]{kitaura2006}.

The canonical progenitor model for an ECSN is a single star that has
evolved to the super-asymptotic giant branch (super-AGB)
\citep[e.g.][]{eldridge2004,siess2007,eldridge2007,poelarends2008,takahashi2013,jones2013,doherty2015}. A
super-AGB star has a hydrogen-rich envelope with a mass of several
\Msun\ around an ONeMg core supported by electron degeneracy
pressure. They are otherwise similar in structure to normal AGB stars
that have a CO core. Because of the hydrogen-rich envelope an ECSN from a
super-AGB star is presumed to be observed as a Type~II SN
\citep{tominaga2013}. However the exact observational nature of
explosions of super-AGB stars is uncertain. They are expected to have
high mass-loss rates in a slow stellar wind which leads to a dense
circumstellar environment around the star that any SN ejecta would
interact with if the star was to explode
\citep[e.g.][]{poelarends2008,woosley2015}. Thus it is possible that
ECSNe from super-AGB stars could show signatures of the interaction
between SN ejecta and dense circumstellar media 
\citep[e.g.][]{smith2013,moriya2014}.

However, recent evidence has begun to show that most massive stars,
that give rise to core-collapse SNe, are in fact in binary systems
\citep{2010ApJS..190....1R,sana2012,sana2013,2013ARA&A..51..269D}. These
studies show that maybe 50 to 70 per cent of massive stars have their
evolution affected by binary interactions. For around the mass range
expected for ECSNe of $\sim 8\Msun$ the binary frequency is at
least 50 per cent. It has been suggested that duplicity of a star will
aid its evolution to an ECSN
\citep{nomoto1985,podsiadlowski2004}. ECSNe from binary systems can
have profoundly different observational properties from those from
single stars. For example, most single ECSN progenitors can retain
their hydrogen-rich envelope up until the time of the collapse of their ONeMg
core and thus be observed as Type~II SNe. This is not necessarily
true for ECSNe from binary systems as binary interactions, either
Roche-lobe overflow or common envelope ejection, can remove some or
all of the hydrogen-rich envelope. Therefore many binary ECSNe will occur
with little or no hydrogen remaining. In addition for some the
remaining helium layer can also be severely depleted by the same
interactions.

In this paper, we investigate observational properties of
stripped-envelope ECSNe based on binary population synthesis models
from the Binary Population and Spectral Synthesis (BPASS)
code\footnote{\url{http://bpass.auckland.ac.nz}} \citep[][Eldridge
  et al. in prep.]{EIT2008,ELT2011,SEB2016}. We take the predicted
stellar properties of ECSN progenitors from BPASS that are the result
of binary evolution and investigate observational properties namely
event rates and bolometric light curves (LCs) of stripped-envelope
ECSNe.

We first introduce the BPASS stellar evolution models and discuss ECSN
rates and stripped-envelope ECSN progenitor properties in
Section~\ref{sec:progenitors}. Then, we discuss expected
stripped-envelope ECSN LC properties in Section~\ref{sec:LCs}. We
further discuss the stripped-envelope ECSN properties and possible
observational candidates in Section~\ref{sec:discussion}. We conclude
this paper in Section~\ref{sec:conclusions}.

\section{Progenitor properties}\label{sec:progenitors}
\subsection{Binary population synthesis model}

\begin{figure*}
 \begin{center}
  \includegraphics[width=0.66\columnwidth]{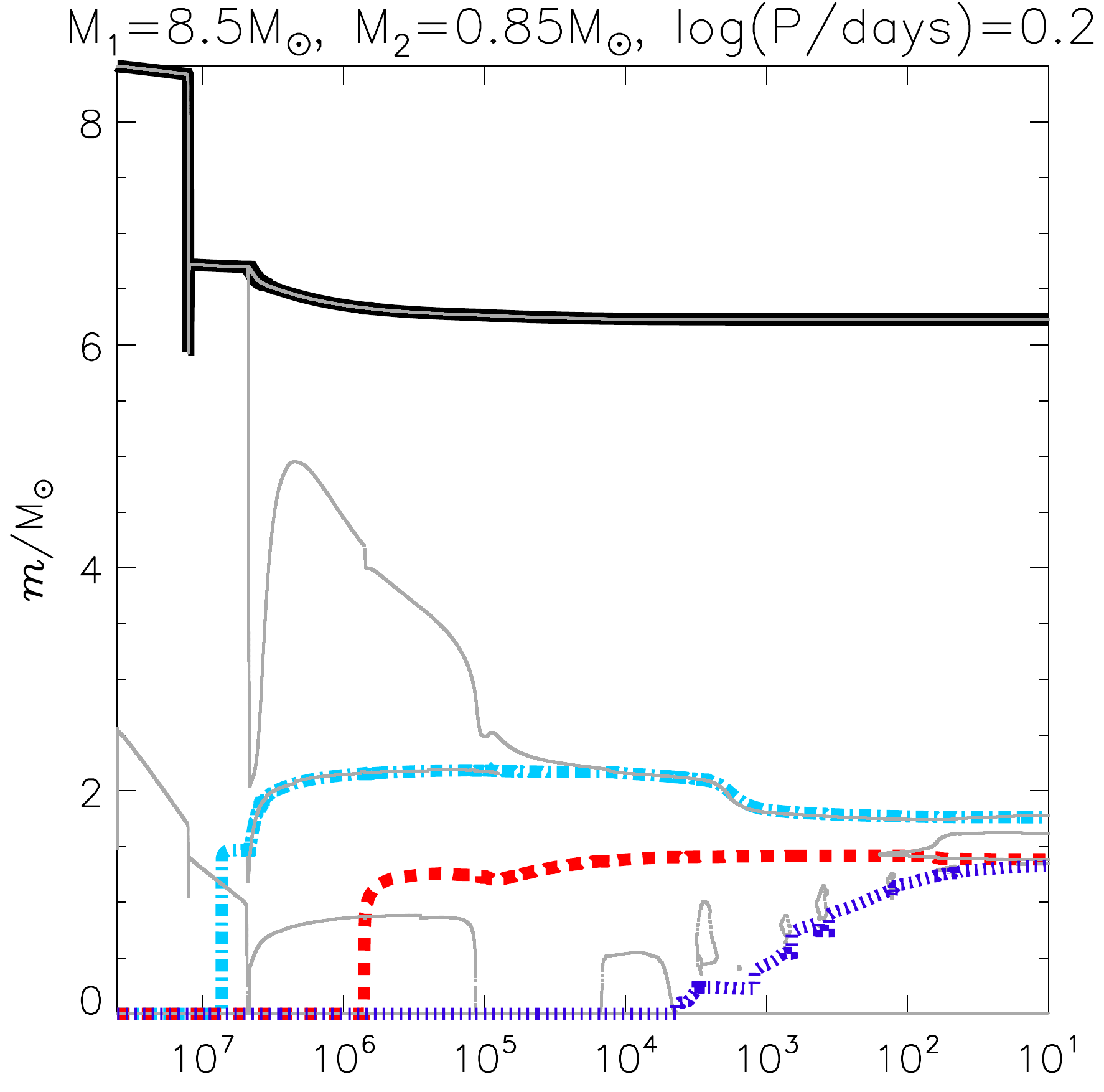}
  \includegraphics[width=0.66\columnwidth]{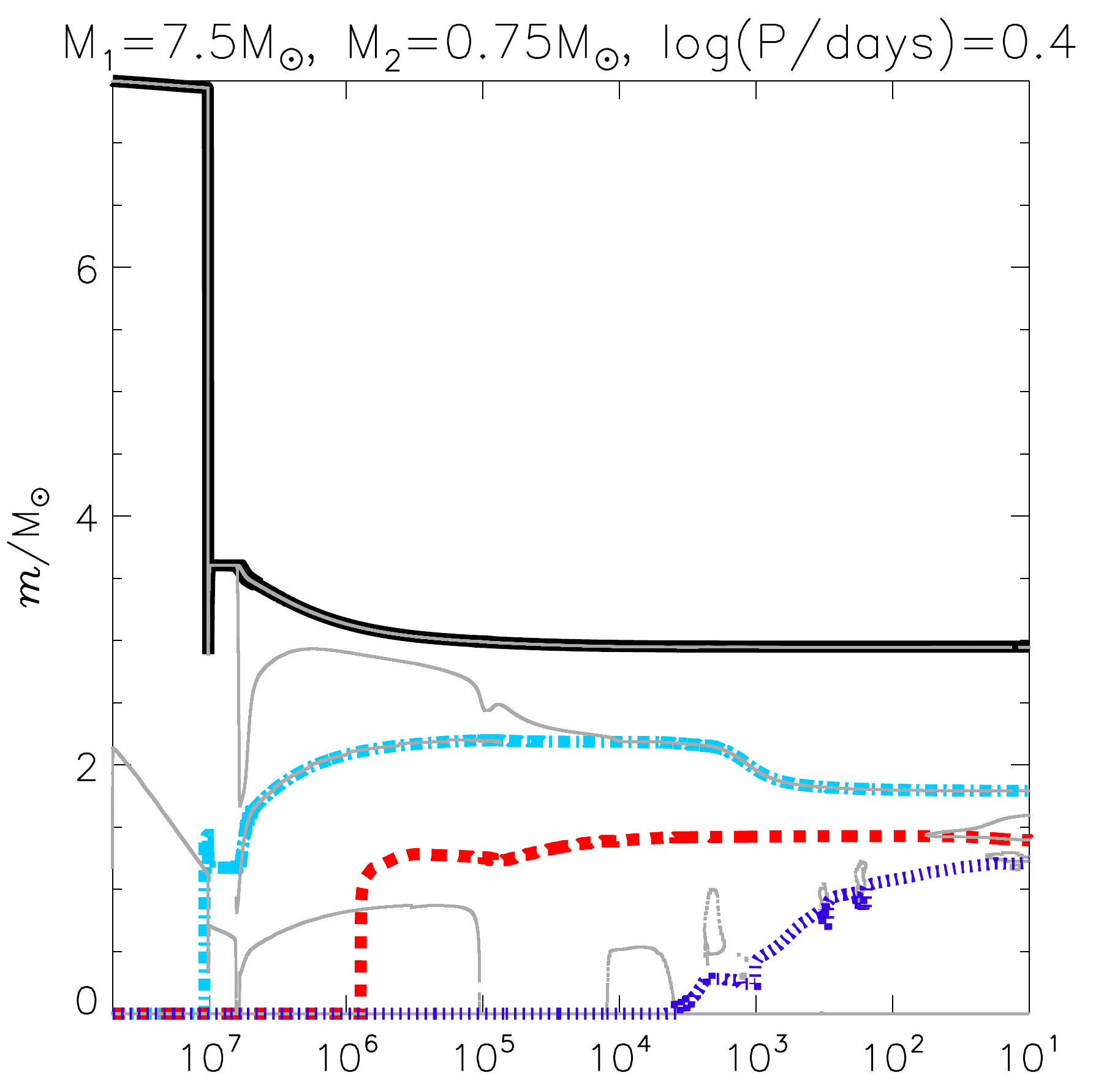} 
  \includegraphics[width=0.66\columnwidth]{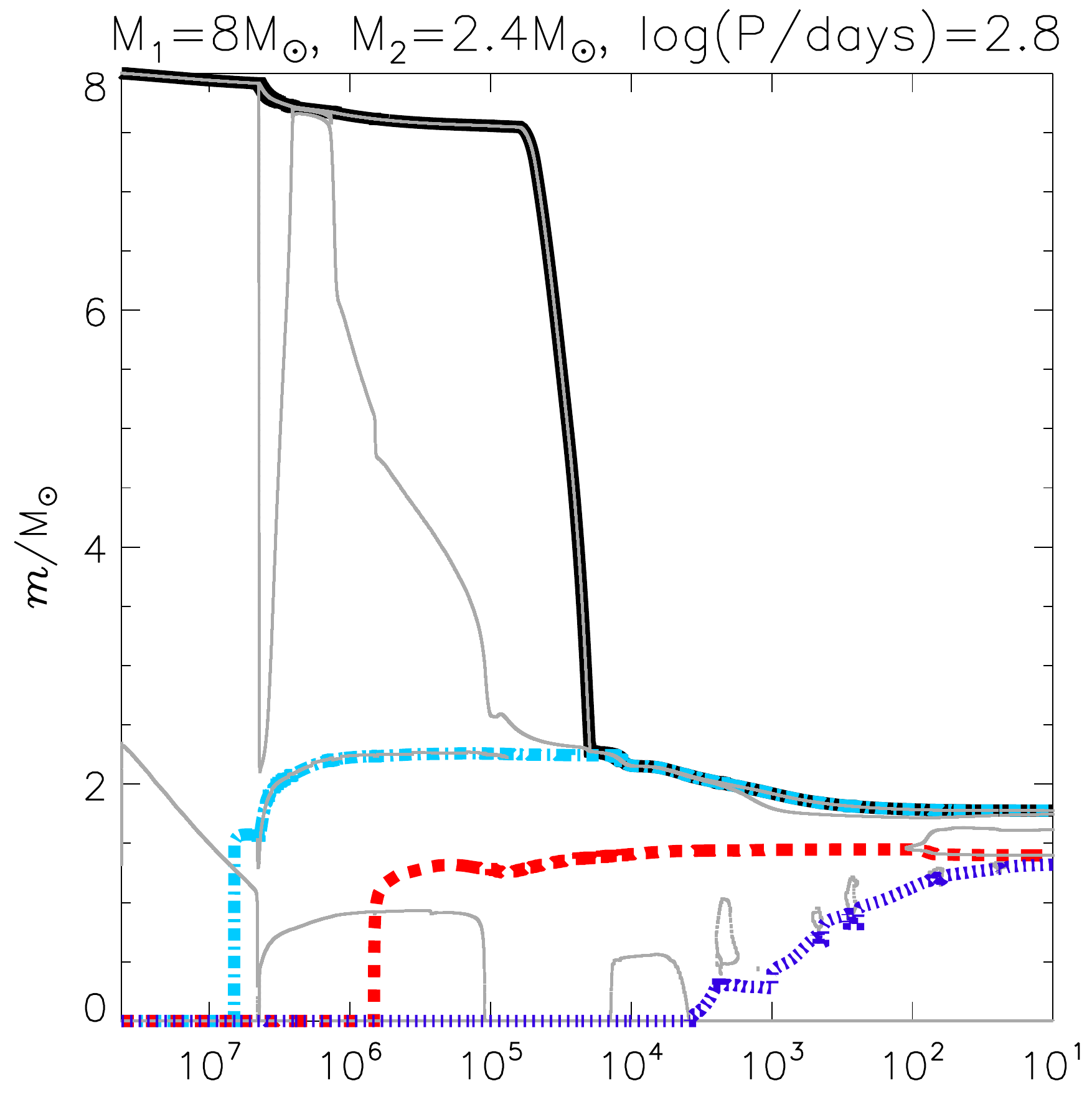} 
  \includegraphics[width=0.66\columnwidth]{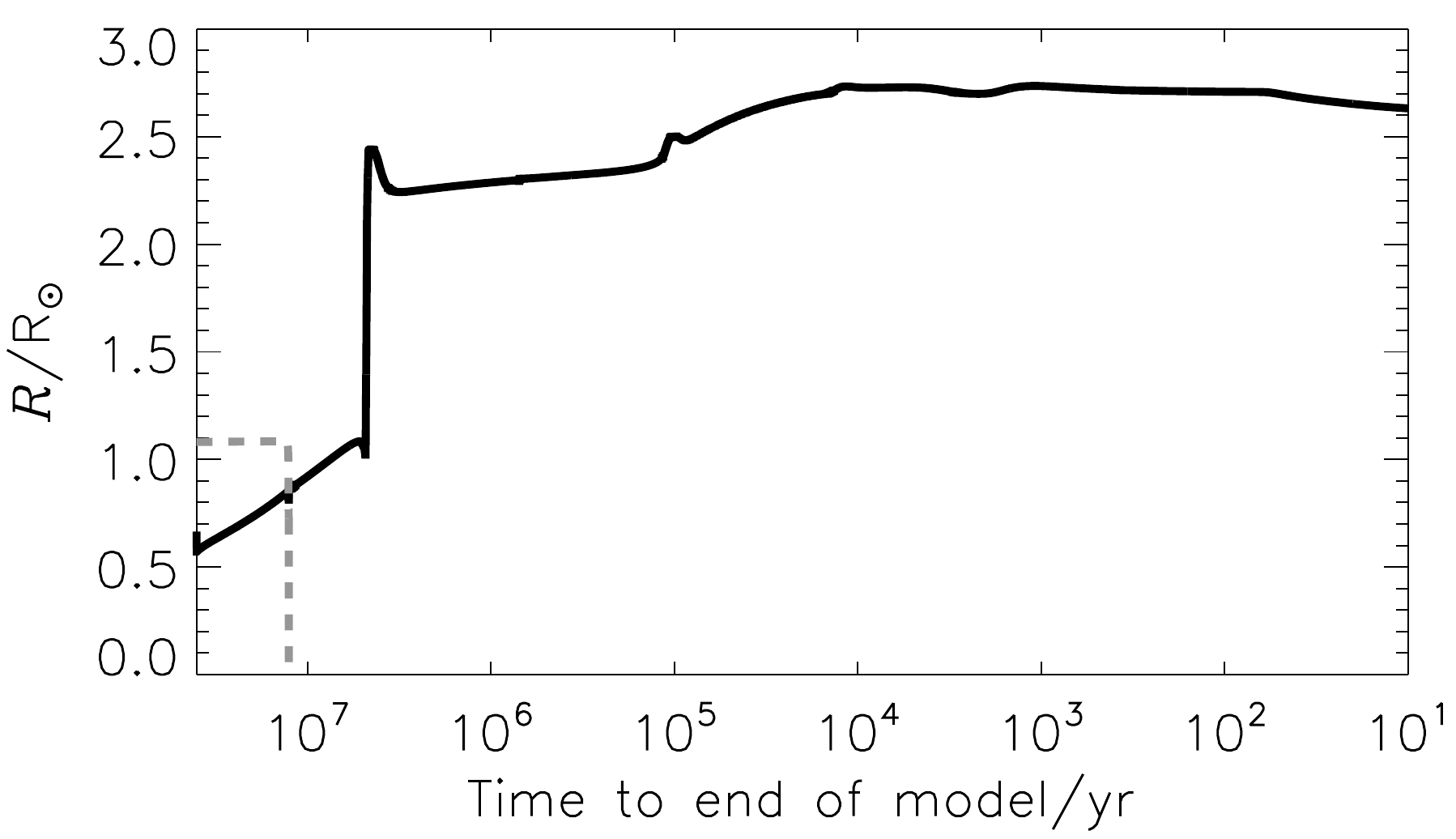}
  \includegraphics[width=0.66\columnwidth]{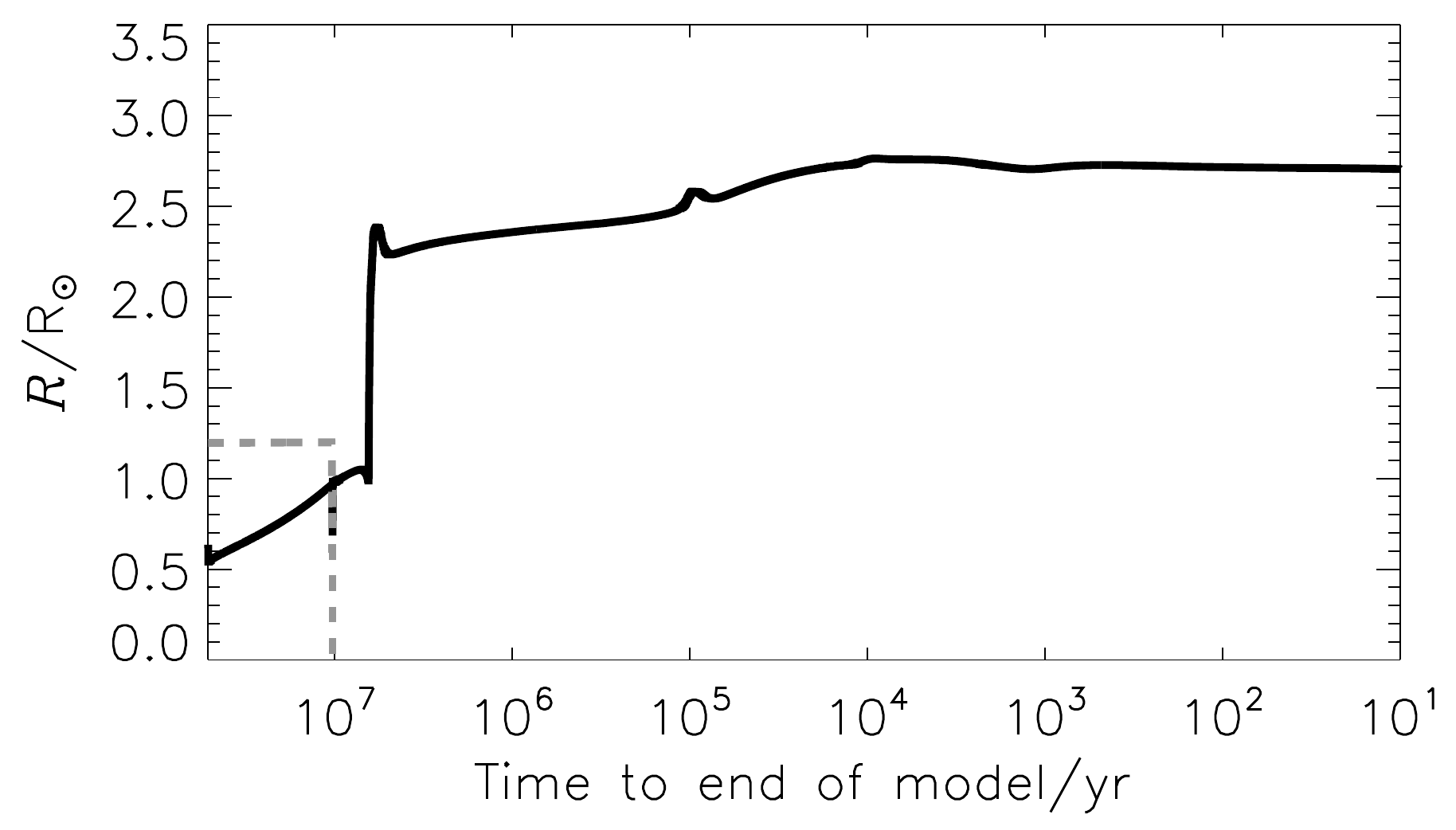} 
  \includegraphics[width=0.66\columnwidth]{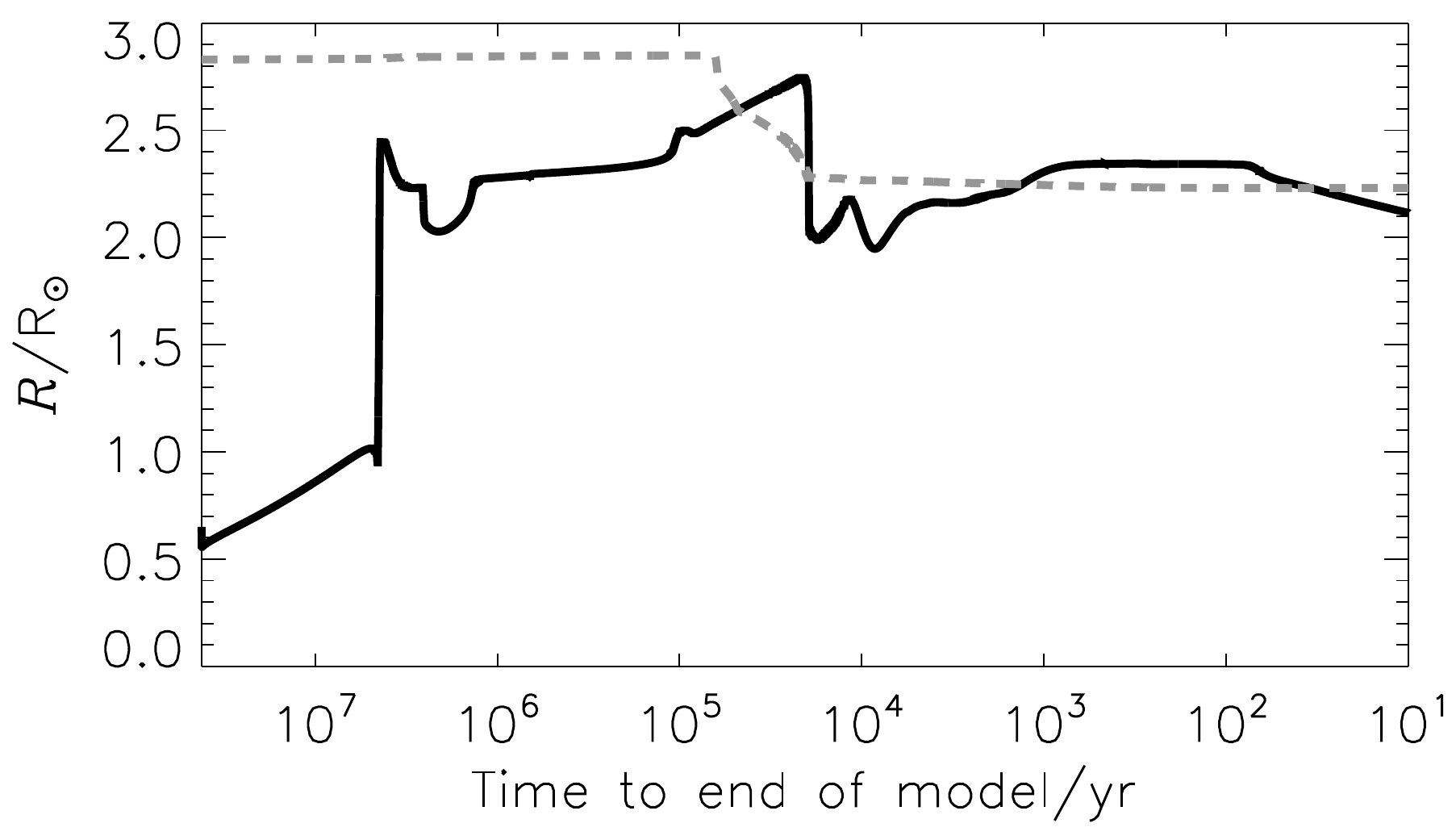} 
 \end{center}
\caption{Example Kippenhahn diagrams for the evolution of ECSN
  progenitors. Each model is at a metallicity of $Z=0.020$. The
  initial mass of the two stars and their initial period are given at
  the top of each panel. In the upper panels the solid black line
  shows the total mass of the star. The light-blue dash-dotted line is the
  edge of the helium core, the red dashed line is the edge of the CO
  core, the blue dotted line is the edge of the ONe core and the thin
  grey lines indicate convective boundaries. In the lower panels the
  black line indicates the radius of the primary star and the grey dashed
  line indicates the binary separation.}\label{fig:kipp}
\end{figure*}

\begin{figure}
 \begin{center}
  \includegraphics[width=\columnwidth]{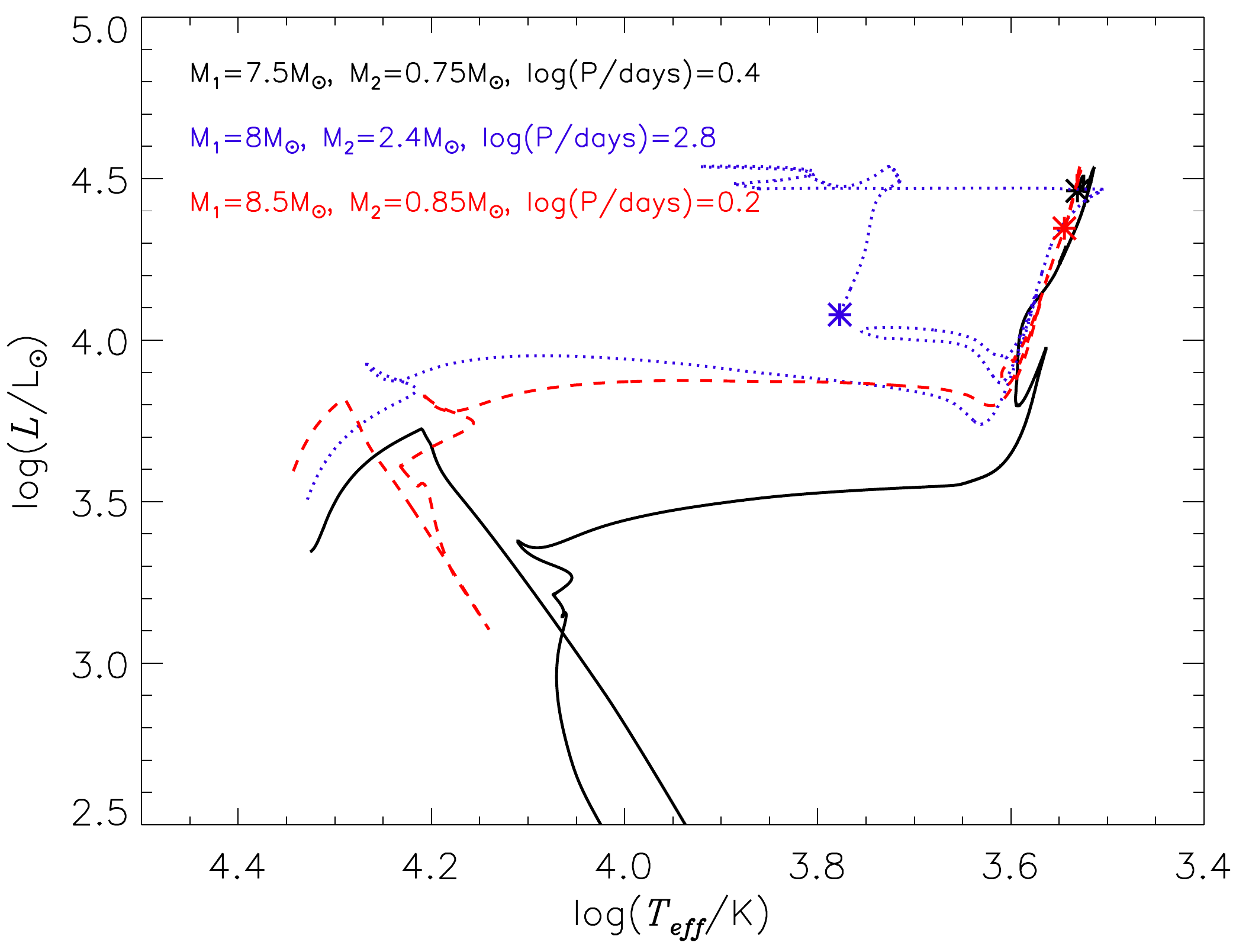} \\
 \end{center}
\caption{The HR diagram showing the evolution of
  the same models as in Figure \ref{fig:kipp}.}\label{fig:hrdiag}
\end{figure}

The stellar evolution models used in this study are from the Binary
Population and Spectral Synthesis, \textsc{BPASS}, code. The model
construction is discussed in detail in \citet{EIT2008} and
\citet{ELT2011}. Here we use the v2.0 models which are almost
identical to those in earlier versions but with an expanded number of
models and a revised treatment for stellar mergers which are described
in \citet{SEB2016} and Eldridge et al. (in prep.). We however present
a brief summary here with the relevant details to this study.

The stellar evolution models are calculated with a modified version of
the Cambridge STARS code as described in \citet{EIT2008}. However the
resolution of the grid used to cover the initial parameters of
metallicity, initial masses and initial period has been increased so
that in total there is a grid of 250,000 detailed stellar models that
allow us to investigate the rich diversity of the evolution of
interacting binary stars. The use of detailed models allows us to
accurately follow how the stellar envelope responds to mass loss --
key to determining the eventual mass and fate of the star. There are
many groups that study binary evolution of massive stars and its
importance for massive stars is review by \citet[][]{vanbev2015}.

We have recently released version 2.0 of BPASS \citep[][Eldridge et
  al. in prep.]{SEB2016}. This incorporates many refinements to the
code and its outputs compared to the earlier versions, mainly in the
spectral synthesis results. The results of BPASS v2.0 have already
demonstrated the improvement in agreement between observations and
stellar population models that arises from the inclusion of
interacting binaries
\citep[e.g. ][]{SEB2016,2016arXiv160103850W,Eld16}.  We use an
initial-mass function based on \citet{1993MNRAS.262..545K}, with a
power-law slope of $-1.3$ between initial masses of 0.1 to
0.5\,M$_{\odot}$ and a slope of $-2.35$ from 0.5 to
300\,M$_{\odot}$. This is combined with an initial-mass ratio of
$q=M_2/M_1$ where we have models with $q=0.1$ to 0.9 in steps of
0.1. We assume that these models are equally weighted. All secondary
stars contribute to the stellar mass but we do not include a companion
in the stellar mass estimate if its initial mass is less than
0.1\,M$_{\odot}$. The initial-period distribution is uniformly
distributed in $\log$ of the period from 1 day to $10^4$ days. There
are some indications that short period systems are more numerous
\citep{sana2012,2013ARA&A..51..269D}, but we retain the uniform
distribution for simplicity. The uncertainty in period distribution
is degenerate with the uncertainty in how we model Roche-lobe
overflow and common-envelope evolution.

Convection is modelled using mixing-length theory and mixing is
modelled simultaneously with the structure with a diffusion equation
that allows us to correctly follow stable semi-convection. Convective
overshooting is modelled by modifying the classical Schwarzschild
criterion for instability with an overshooting parameter of
$\delta_{\rm ov}=0.12$, a value derived from observations of binary
systems by \citet{1997MNRAS.285..696S} and
\citet{1997MNRAS.289..869P}. The mass-loss scheme is discussed in
detail in \citet{EIT2008}. We use the rates of \citet{Vinketal} for
main-sequence OB stars, the rates of \citet{2000A&A...360..227N} for
Wolf-Rayet stars and those of \citet{1988A&AS...72..259D} for all
other stars.  We scale the mass-loss rates applied from those observed
in the local universe, such that
$\dot{M}(Z)=\dot{M}(Z_{\odot})(Z/Z_{\odot})^{\alpha}$ and $\alpha=0.5$
\citep[except in the case of OB stars where $\alpha = 0.69$,
  see][]{Vinketal}. There is little consensus in the literature
regarding the definition of solar metallicity.
\citet{2014ApJ...787...13V}, for example, suggest the metal fraction
in the Sun is rather higher than usually assumed, while some authors
\citep{2002ApJ...573L.137A,2005ARAA..43..481A} suggest that Solar
metal abundances should be revised downwards to closer to Z = 0.014
\citep[also appropriate for massive stars within 500pc of the
  Sun,][]{2012AA...539A.143N}. We retain $Z_{\odot}=0.020$ for
consistency with the empirical mass-loss rates which were originally
scaled from this value. In this study we also use models with a
metallicity of $Z=0.004$ for low metallicity environments like the
Small Magellanic Cloud.

Our models do not include any rotational mixing, although at low
metallicities we do include an approximate implementation of
quasi-chemically homogeneous evolution due to rapid rotation
\citep{ELT2011,SEB2016}. This does not occur in the initial mass range
of the ECSN progenitors as we only include this for stars with an
effective initial mass greater than 20M$_{\odot}$. We do not include
wind-accretion in our binary models. Roche-lobe overflow is assumed to
occur when the primary star filled its Roche-lobe radius and the
mass-loss rate is dependent on how far it overfills the Roche-lobe as
described in detail in \citet{EIT2008}. If the mass loss does not
prevent the star from expanding and the star grows to engulf its
companion, common-envelope evolution is assumed to occur. In our
code, it is numerically difficult to remove the stellar
envelope instantaneously. We therefore set a high mass-loss rate and
relate the decrease in the binary orbital energy to the amount of
binding energy required to lose material from the surface of the
primary star as described in \citet{EIT2008}. One difference to our
model now is that if the secondary star also fills its Roche-lobe
during common-envelope evolution the stars are assumed to merge and
the mass of the secondary is added to the remaining mass of the
primary star.

From our grid of stellar models we select out ECSN progenitors using
the mass ranges of \citet{woosley2015} for the CO core mass. If a
star has experienced core carbon burning and has a CO core mass between
1.38 and 1.4~\Msun, it does not
experience further burning and collapses due to being above the
Chandrasekhar mass. However, whether the star experiences an
electron-capture SN or an iron-core collapse is uncertain and depends on
how details such as the Urca process are modelled
\citep{2014ApJ...797...83J}. We assume CO cores with masses above
1.4M$_{\odot}$ do continue growing eventually to iron-core collapse
\citep{woosley2015}. Stars below 1.38~\Msun typically require the core
to grow either by helium shell burning or thermal pulses before they
could collapse. We assume if this is the case that during the time for
the core to grow stellar wind mass loss removes the envelope and the
star forms a white-dwarf rather than an ECSN.

We note that we also require a total stellar mass greater than
1.5~\Msun\ for an ECSN to occur. Only one of our models has a massive
enough CO core but is below this mass limit.  Such a star below our
assumed mass limit would have very small ejecta masses and
correspondingly shorter timescales of their LC evolution. The
LC also becomes dependent on exactly where we chose to put in
a mass cut for the amount of material ejected as shown by
\citet{tauris2013}.

We show sample evolution of some of our progenitors in Kippenhahn and
Hertzsprung-Russell (HR) diagrams in Figures \ref{fig:kipp} and
\ref{fig:hrdiag}. All three example cases are at $Z=0.020$. The first
two cases are for stellar mergers that occur after a short common-envelope
phase during the main-sequence. In the second case the merger
happens towards the end of the main-sequence and so the helium core of
the star is larger than what would be expected for a similar star of
the same initial mass. In the third case the common-envelope evolution
occurs after core helium burning (Case C) and so their hydrogen-rich
envelope is lost leaving a helium star. This star eventually becomes a
helium giant which experiences another interaction reducing the helium
envelope further. In all three cases we see that carbon-burning
ignites centrally and burns outwards in a series of flashes. Also
towards the end of the evolution we see that a convection zone opens
up at the edge of the CO core. This reduces the CO core mass to below
1.4M$_{\odot}$ and prevents the ONe core from growing further. The
formation of this convection zone reduces the core mass into our
assumed range for ECSNe. At this point our code cannot calculate models
further. The few models that we have been able to take further find
that the central density increases and conditions for an ECSN can be
obtained.

We stress here that while we use a detailed evolution code the ethos
behind out models is to have greater accuracy in our models than using
rapid stellar evolution such as that from,
e.g. \citet[][]{2000MNRAS.315..543H}, but we cannot calculate large
numbers of detailed models over a broad initial parameter space and
include all physical effects such as the models of
\citet{2014ApJ...797...83J,tauris2015} which is why we need to make
assumptions about which of our models will experience ECSNe.  We note
that our helium star models are similar to those recently discussed by
\citet{tauris2015}. Therefore while our evolutionary models are not
evolved until the point of core collapse their similarity to these
models suggests that they will do so. An advantage of our models 
is that we include ECSNe from both primary and secondary star as well
as via mergers, by following all possible channels, allowing us to
estimate the rates of these events.

The channels for the ECSNe are similar at both metallicities as shown
in Figure \ref{fig:kipp}. The initial mass range of the progenitors at
$Z=0.020$ ranges from 4 to 12~\Msun. The two channels are either Case
A mass-transfer that normally leads to a stellar merger. When mergers
occur earlier in the main-sequence the final mass of the progenitor is
similar to that expected from single-star ECSN evolution. Later Case A
mergers tend to have less massive hydrogen-rich envelopes at the time of
core-collapse. The other main channels are later Case B or Case C
interactions which avoid a merger and remove the hydrogen-rich envelopes
leaving helium stars as described by \citet{podsiadlowski2004}. Due to
the low mass of the helium stars these tend to become helium giants
and some experience a second binary interaction that reduces the
helium envelope mass. The nature of the progenitors when they explode
in both cases are relatively cool and luminous progenitors as those
with hydrogen-rich envelopes are cooler than the helium progenitors.

\subsection{Stripped-envelope ECSN progenitor properties}

Table~\ref{table:ecsnprop} summarizes predicted rates of ECSNe from
BPASS. The total ECSN fractions in core-collapse SNe are 1.5 per cent
for $Z=0.020$ and 8.6 per cent for $Z=0.004$. This increase is due to
less mass loss because of weaker stellar wind at the lower
metallicity.

We further show expected SN-type fractions of ECSNe in
Table~\ref{table:ecsnprop}. We classify ECSNe from progenitors with a
remaining hydrogen mass less than 0.5~\Msun\ as stripped-envelope
ECSNe (Type~IIb and Type~I) and the others as Type~II.  Type~IIb SNe
are typically estimated to have hydrogen masses of $0.1-0.5~\Msun$
\citep[e.g.][]{woosley1994} and we choose 0.5~\Msun\ as the dividing
mass. We find in our models that few models lie close to this mass
with most have a few \Msun of hydrogen or no hydrogen. We find that 51
per cent of ECSN progenitors at $Z=0.020$ are stripped-envelope ECSNe
and the fraction reduces to 14 per cent at $Z=0.004$. However, the
stripped-envelope ECSN fractions in core-collapse SNe are slight
higher at $Z=0.004$ (1.2 per cent) than at $Z=0.020$ (0.8 per
cent). This suggests that at lower metallicity there more
Type~II ECSNe.

\begin{table}
 \caption{The relative fraction for ECSNe as estimated from BPASS.
 Statistical errors are presented.}
 \label{table:ecsnprop}
 \begin{tabular}{ccccc}
  \hline
  $Z^a$ & ECSN fraction$^b$ & \multicolumn{3}{c}{Type distribution in ECSNe} \\
    &  & Type~I & Type~IIb & Type~II \\
  \hline
  $0.020$ & $0.015\pm0.002$ & $0.42\pm0.09$ & $0.09\pm0.05$  & $0.49\pm0.12$ \\
  $0.004$ & $0.086\pm0.006$ & $0.11\pm0.02$ & $0.03\pm0.01$ & $0.86\pm0.12$ \\
  \hline
\multicolumn{5}{l}{$^a$ Initial progenitor metallicity.}\\
\multicolumn{5}{l}{$^b$ Fractions of ECSNe in core-collapse SNe.}\\
 \end{tabular}
\end{table}

\begin{figure}
 \begin{center}
  \includegraphics[width=\columnwidth]{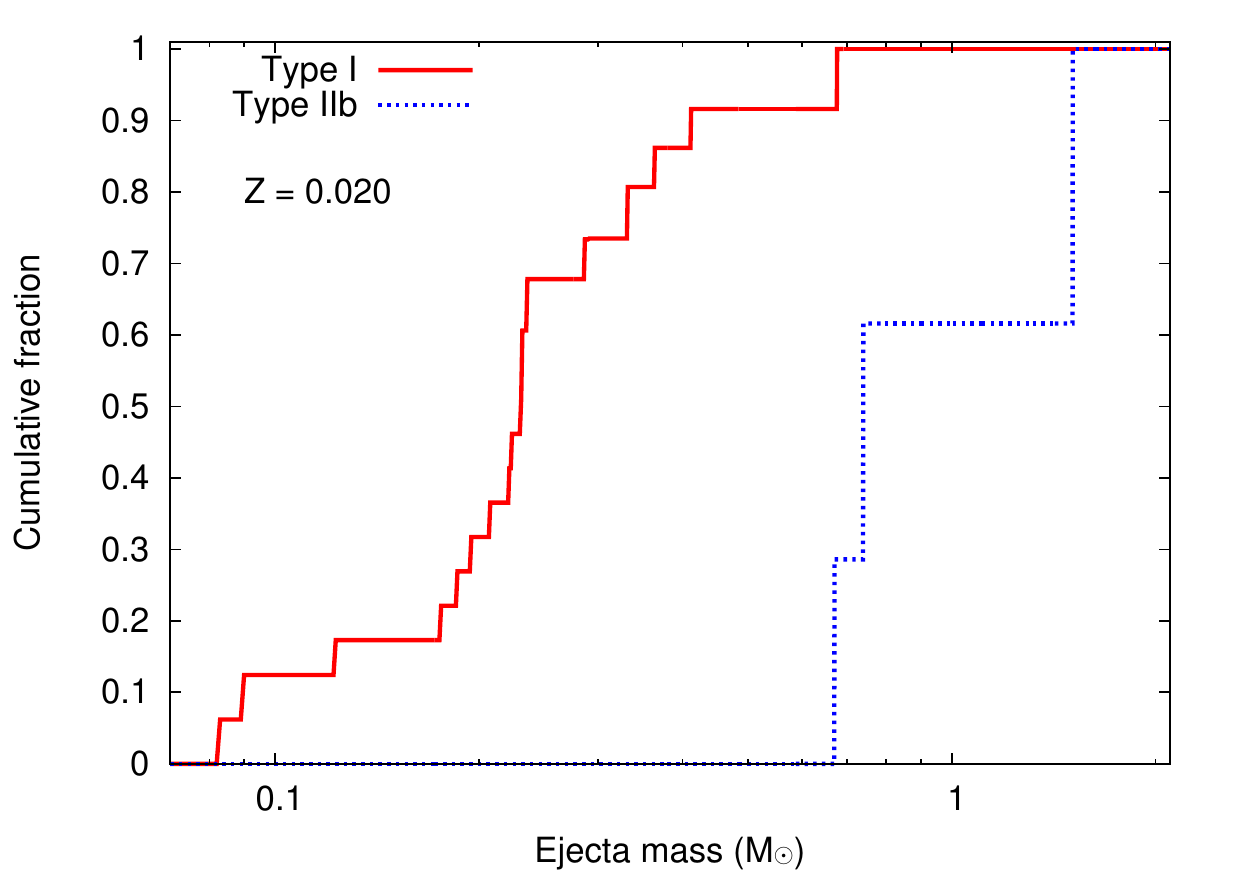} \\
  \includegraphics[width=\columnwidth]{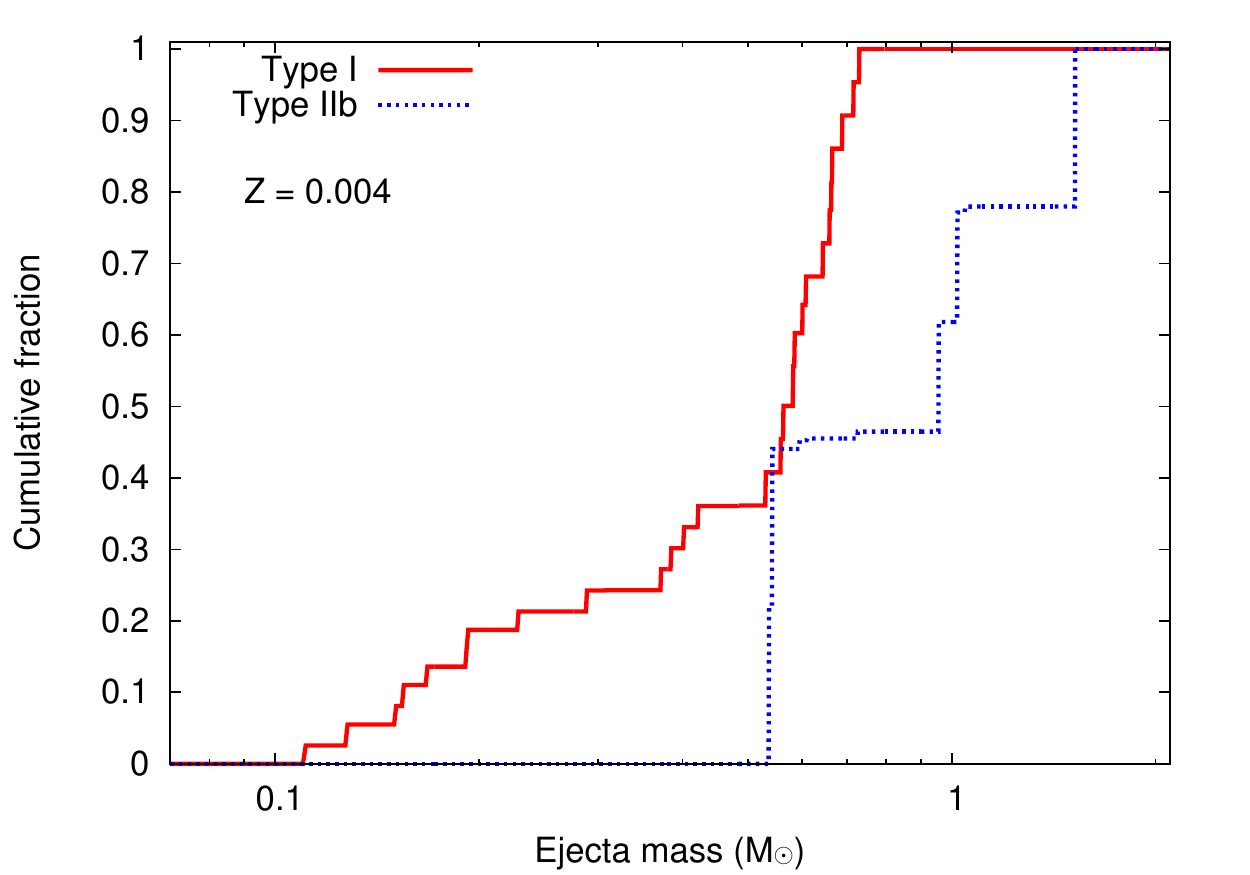} 
 \end{center}
\caption{ The cumulative ejecta mass distribution for the ECSNe taken from
  BPASS. The upper panel is for a metallicity of $Z=0.020$ and the
  lower panel is for a metallicity of $Z=0.004$.
}\label{fig:massdistribution}
\end{figure}

\begin{figure}
 \begin{center}
  \includegraphics[width=\columnwidth]{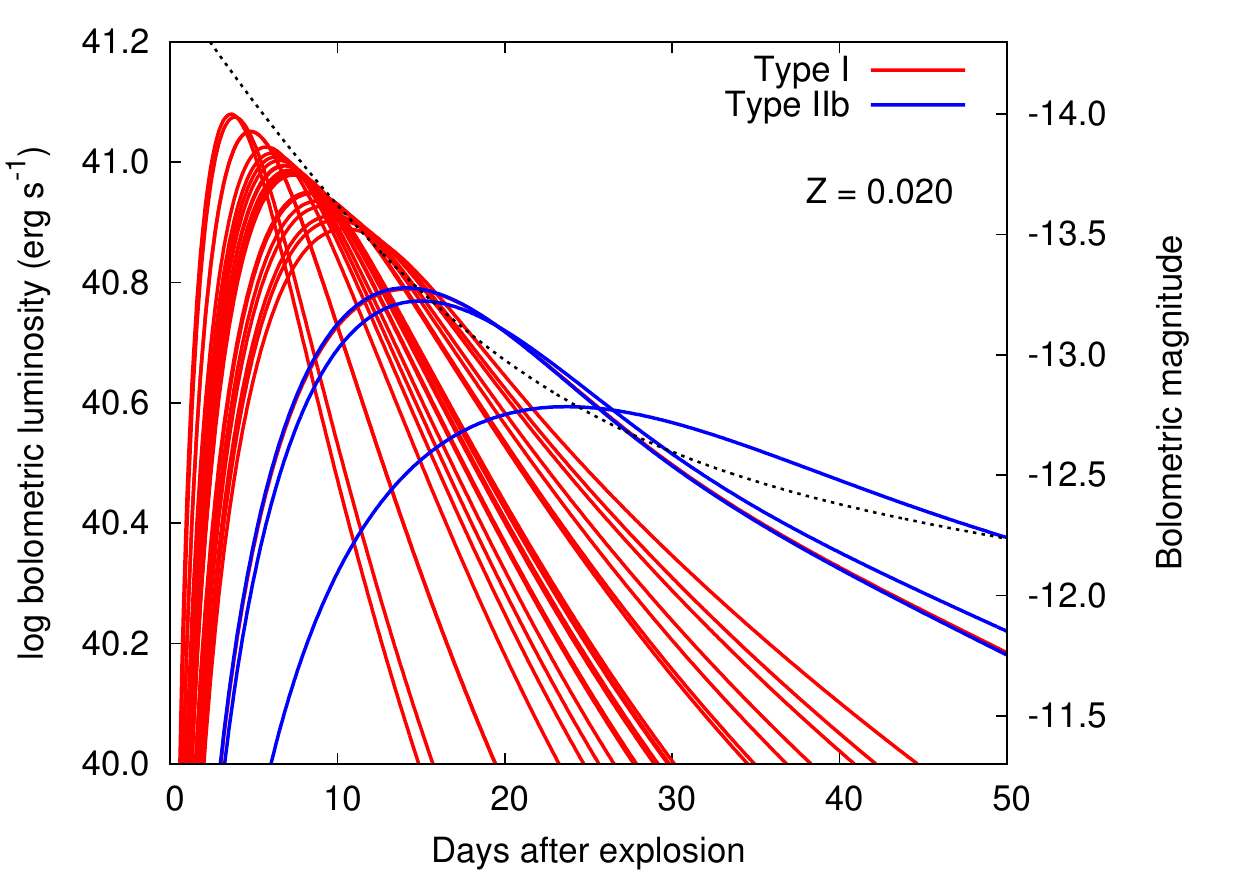} \\
  \includegraphics[width=\columnwidth]{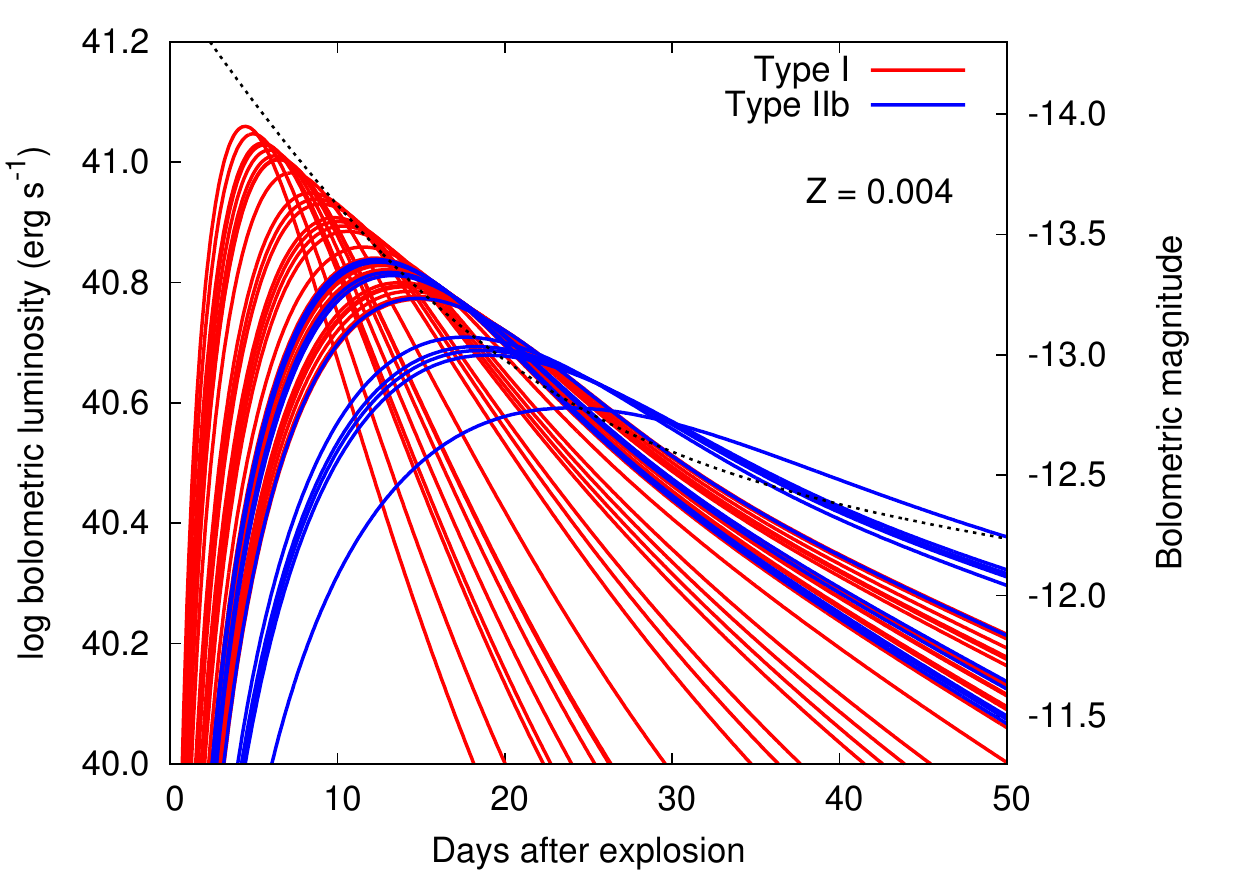} 
 \end{center}
\caption{ Bolometric LCs of stripped-envelope ECSNe.
  We show LCs of all the ECSN progenitors from BPASS and
  each line represents a bolometric LC from a progenitor.
  The dotted lines are the total available nuclear energy from $2.5\times
  10^{-3}~\Msun$ of \Ni. The upper
  panel is for a metallicity of $Z=0.020$ and the lower panel is for
  $Z=0.004$.}\label{fig:lightcurves}
\end{figure}

We divide stripped-envelope ECSNe into two different subtypes,
Type~IIb and Type~I. Those with hydrogen mass between 0.5~\Msun\ and
0.01~\Msun\ are classified as Type~IIb and the others as Type~I because even small amounts of hydrogen can be observed in SN spectra
\citep[e.g.][]{dessart2011,hachinger2012}.  The majority of
stripped-envelope ECSNe are Type~I. We do not sub-classify Type~I
here.  Whether Type~I SNe are classified as Type~Ib or Ic depends on
the degree of \Ni\ mixing \citep[e.g.][]{dessart2012}. All the Type~I
ECSNe we obtain have helium masses of more than 0.06~\Msun, which is
the maximum amount of helium that can be hidden
\citep{hachinger2012}. Thus, if sufficient \Ni\ mixing occurs, Type~I
ECSNe may be observed as Type~Ib SNe.

Because the explosion energy (\Eej) and the \Ni\ mass (\Mni) are
expected to vary little among ECSNe, the ejecta mass (\Mej) is the
most influential parameter determining the LC properties of
stripped-envelope ECSNe. Figure~\ref{fig:massdistribution} shows the
ejecta mass distributions of stripped-envelope ECSNe from BPASS. A
typical ejecta mass of stripped-envelope ECSNe is expected to be of
the order of 0.1~\Msun. The average ejecta mass increases as the
metallicity decreases because of the the scaling of stellar winds by
initial metallicity as described above.

\begin{figure}
 \begin{center}
  \includegraphics[width=\columnwidth]{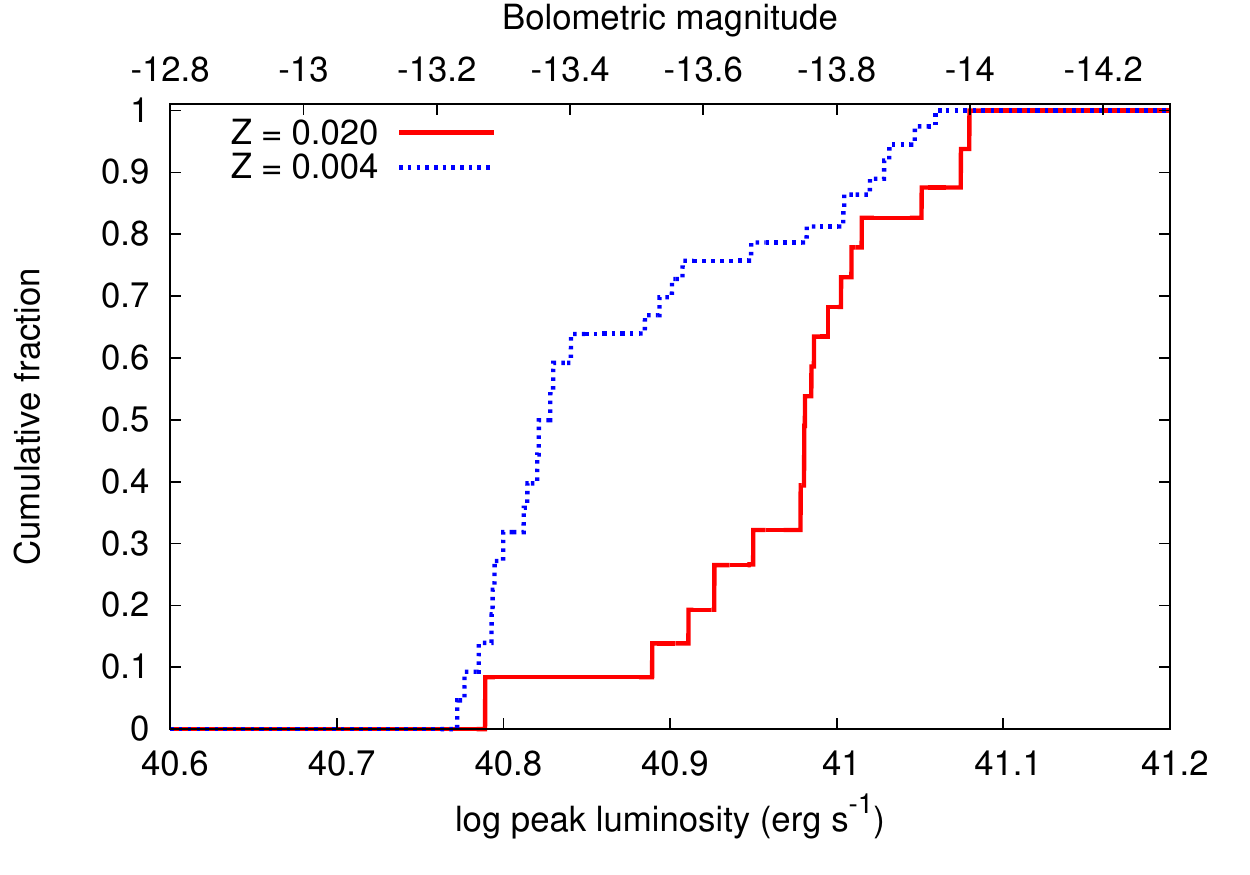}
 \end{center}
 \begin{center}
  \includegraphics[width=\columnwidth]{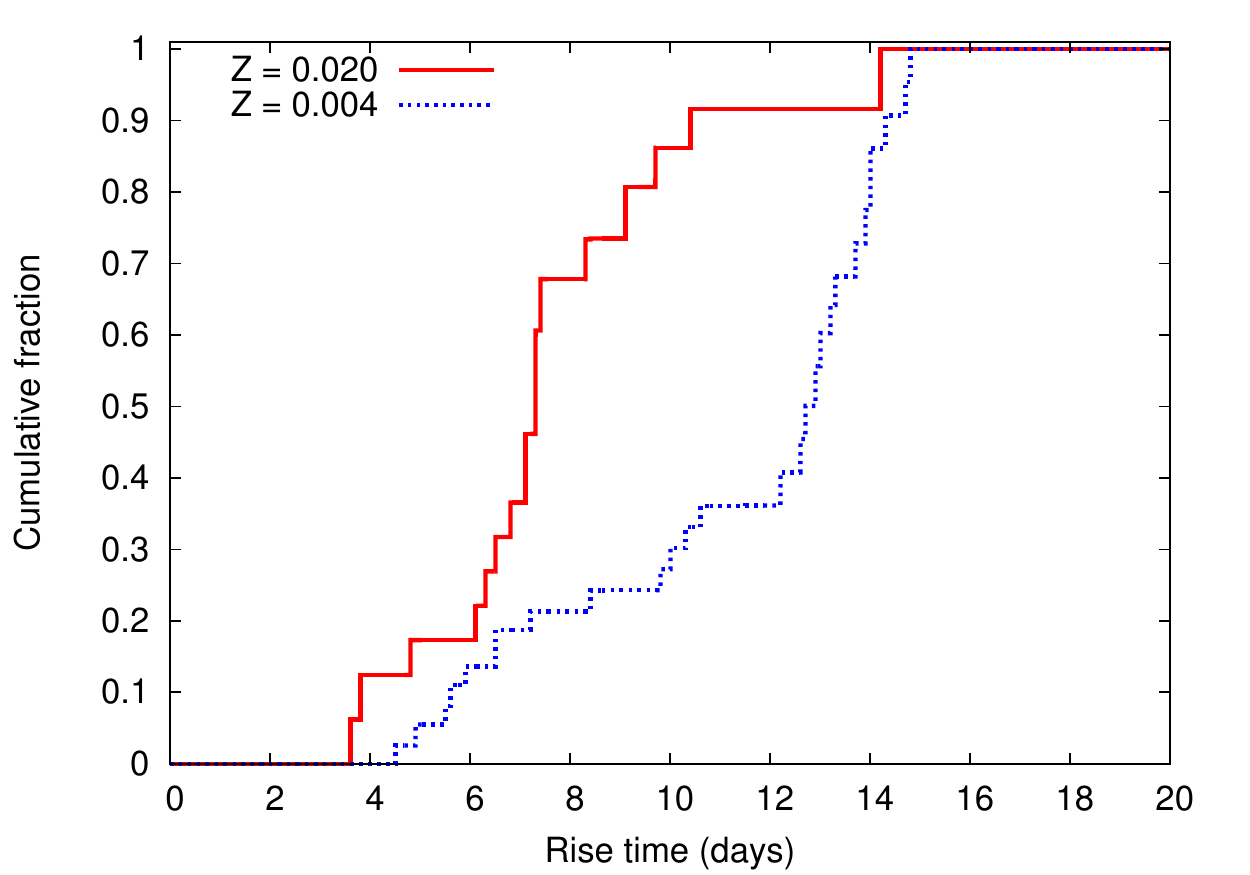}
 \end{center}
\caption{The upper panel shows the cumulative peak luminosity distribution
  of stripped-envelope ECSNe. The
  lower panel shows the cumulative rise time distributions of stripped-envelope
  ECSNe. }\label{fig:fractions}
\end{figure}

\section{Stripped-envelope ECSN light curves}\label{sec:LCs}
\subsection{Method}
We calculate bolometric LCs of stripped-envelope ECSNe by using a
simplified analytic method introduced by \citet{arnett1982}. This
simplified method has been used to obtain bolometric LCs of
stripped-envelope SNe where the effect of hydrogen recombination in
opacity is negligible
\citep[e.g.][]{valenti2008,chatzopoulos2012,inserra2013}. Briefly, we
numerically evaluate the following equation to obtain the bolometric
luminosity $L$ at time $t$ after the explosion,
\begin{equation}
L(t)=\int^{\frac{t}{\tau_m}}_02\tau_m^{-2}DL_\mathrm{decay}(t')t'e^{\left(\frac{t'-t}{\tau_m}\right)^2}dt',
\end{equation}
where $\tau_m$ is the effective diffusion time, $L_\mathrm{decay}(t)$
is the energy input from the nuclear decay of
$\Ni\rightarrow\Co\rightarrow\Fe$, and $D$ is the deposition function
\citep{arnett1982}. We only consider gamma-rays from the nuclear decay
as a heating source in SN ejecta. Assuming the homogeneous density
distribution, $\tau_m$ can be expressed as
\begin{equation}
\tau_m = \left(\frac{6}{5}\right)^{0.25}\left(\frac{\kappa_e}{\beta c}\right)^{0.5}\Mej^{0.75}\Eej^{-0.25},
\end{equation}
where $\kappa_e$ is an electron-scattering opacity in SN ejecta, $c$
is the speed of light, and $\beta \simeq 13.8$ \citep{arnett1982}. The
electron-scattering opacity is assumed to be
$0.1~\mathrm{cm^2~g^{-1}}$. The deposition function $D$ approximately
takes the gamma-ray trapping efficiency in SN ejecta into account. It
is evaluated as in \citet{arnett1982} by assuming the gamma-ray
opacity of $0.03~\mathrm{cm^2~g^{-1}}$ \citep[e.g.][]{colgate1980}.

Both one- and two-dimensional numerical explosion simulations of ECSNe
consistently show that their explosion energy and \Ni\ mass are $\sim
10^{50}~\mathrm{erg}$ and $\sim 10^{-3}~\Msun$, respectively
\citep{kitaura2006,burrows2007,wanajo2009,wanajo2011}. Thus, the LC
diversity of stripped-envelope ECSNe mostly come from the diversity in
\Mej\ (Fig.~\ref{fig:massdistribution}). We take \Mej\ from the
population synthesis model and fix $\Eej=1.5\times
10^{50}~\mathrm{erg}$ and $\Mni=2.5\times 10^{-3}~\Msun$ in our LC
calculations \citep{wanajo2011}.

\subsection{Results}
Figure~\ref{fig:lightcurves} presents bolometric LCs of
stripped-envelope ECSNe.  Each line presents a LC from a progenitor obtained from BPASS.
However, the likelihood to obtain each progenitor is not taken into account
in Figure~\ref{fig:lightcurves}.
Figure~\ref{fig:fractions} shows the
expected distributions of their peak luminosity and rise time in which
the probability distribution from the BPASS model is taken into
account.

The typical rise time of stripped-envelope ECSN LCs is about 7~days
($Z=0.020$) and 13~days ($Z=0.004$) which are much shorter than
typical stripped-envelope SNe
\citep[e.g.][]{drout2011,taddia2015,lyman2016}. The rise time is
generally shorter in the higher metallicity models because of the
smaller ejecta mass (Figure~\ref{fig:massdistribution}). The peak
luminosity of stripped-envelope ECSNe is very faint because of their
small \Ni\ mass.  Their typical peak luminosity is expected to be
around $10^{41}~\mathrm{erg~s^{-1}}$ ($-13.8$~mag, $Z=0.020$) or
$7\times 10^{40}~\mathrm{erg~s^{-1}}$ ($-13.4$~mag, $Z=0.004$).  After
reaching the peak luminosity, the LCs decline very rapidly with a
similar timescale to their rise.  The luminosity does reach the
possible maximum from nuclear energy deposition because the small ECSN
ejecta masses cause inefficient gamma-ray trapping and gamma-rays from
the nuclear decay cannot heat the ejecta to a great amount.

There are several caveats in our simplified LC model. First, we have
fixed \Eej\ in our models which can mainly affect the rise
time. However, as the rise time is roughly proportional to $\kappa_e
(\Mej^3/\Eej)^{1/4}$ \citep{arnett1982}, a small variation in
\Eej\ has little effect on the rise time compared to \Mej. We also
fixed \Mni\ which mainly changes the peak luminosity. If the \Mni can
vary by the face of a few, then the LC will also vary by the
same factor (about 1~mag). Even then stripped-envelope ECSNe will
still be very faint transients. We have also fixed the
electron-scattering opacity to $0.1~\mathrm{cm^2~g^{-1}}$, although
this is likely a good approximation \citep[e.g.][]{inserra2013}. The
electron-scattering opacity may vary especially in those progenitors
with a relatively large hydrogen mass (blue LCs in
Figure~\ref{fig:lightcurves}). Overall, although there are several
uncertainties in our LC model due to its simplicity, we expect that
general LC properties we estimate in this section remain unchanged
even if we were to increase the complexity of our models.

\section{Discussion}\label{sec:discussion}
The faintest kind of Type~I SNe currently known are SN~2008ha-like
events which have peak luminosity of around $-14$~mag and rise times
of around 10~days (e.g. SN~2008ha,
\citealt{foley2009,foley2010,foley2016,valenti2009}; SN~2010ae,
\citealt{stritzinger2014}). SN~2008ha-like SNe are often regarded as a
sub-class of SN~2002cx-like (Type~Iax) SNe \citep{foley2013} and
related to partial disruption of white dwarfs
\citep[e.g.][]{kromer2015}.  We find that bolometric LCs of
stripped-envelope ECSNe are consistent with those of SN~2008ha-like
SNe. Assuming typical ejecta masses of 0.3~\Msun\ ($Z=0.020$) and
0.6~\Msun\ ($Z=0.004$), we can estimate the typical ejecta velocities
would be 7000~\kmps\ ($Z=0.020$) and 5000~\kmps\ ($Z=0.004$). The
small photospheric velocities of SN~2008ha-like SNe are consistent
with these estimates, particularly for the low metallicity model
progenitors.

The observed SN of this group that best matches our progenitor models
is SN~2010ae. The estimated explosion properties for SN~2010ae were,
$\Eej\sim 10^{50}~\mathrm{erg}$, $\Mej\sim 0.1~\Msun$, and $\Mni\sim
10^{-3}~\Msun$ \citep{stritzinger2014}. These match our lowest ejecta
mass stripped-envelope ECSN models.

Spectra of SN~2008ha-like SNe are characterized by features of
intermediate-mass elements
\citep{foley2009,foley2010,stritzinger2014}. In principle, spectral
features from intermediate-mass elements can be observed in explosions
of ONeMg cores. Enhanced production of Ca, which has been observed in
SN~2008ha-like SNe, is also predicted \citep{wanajo2013}.  Currently,
spectral modelling of stripped-envelope ECSNe is lacking. We need to
investigate spectral properties of stripped-envelope ECSNe and make
proper comparisons with SN~2008ha-like SNe and other faint and rapid
transients.

It is worth noting that while no progenitors of these events have
been observed, a possible red source at the position of SN~2008ha
found in a late-time image may be a companion star of an ECSN
progenitor \citep{foley2014}. A detailed comparison to stellar models
such as in \citet{Eld16} is beyond the scope of this paper. We do find
some of our models do transfer a large amount of mass to the companion
star which would acclerate its evolution and so it could be observed
as a cool red star after the explosion of the progenitor star.

Both SN~2008ha and SN~2010ae are found in low metallicity environments
which are close to that of the Large Magellanic Cloud
\citep{foley2009,stritzinger2014}. They also appear in star-forming
galaxies and have short delay time \citep{foley2014}. We find that
stripped-envelope ECSN rates are higher in lower metallicity
environments, and this can explain the possible preference of
SN~2008ha-like SNe appearing in low metallicity
environments\footnote{There are only a few SN~2008ha-like
  SNe found so far and this preference has not yet been confirmed
  statistically.}.  Because we expect many more Type~II ECSNe in lower
metallicity environments (Table~\ref{table:ecsnprop}), Type~II ECSNe
have also likely been detected already if SN~2008ha-like SNe are
stripped-envelope ECSNe. It is interesting to note that Type~IIn-P SNe
which are suggested to be ECSNe
\citep[e.g.][]{kankare2012,mauerhan2013} may also prefer lower
metallicity environments \citep{taddia2015a}, and they may correspond
to Type~II ECSNe we predict to exist.

Although SN~2002cx-like SNe share some common properties, they are
also diverse \citep{foley2013}. Most of them are likely related to
explosions or eruptions of white dwarfs, but some of them, especially
the faintest ones like SN~2008ha, may be related to ECSNe as we
suggest here.  It is also interesting to note that \citet{jones2016}
recently argue that oxygen deflagration in ONeMg cores may actually
result in partial disruption of the ONeMg cores.  In addition, massive
stars other than ECSN progenitors can also end up with
rapidly-evolving faint transients
\citep[e.g.][]{tauris2013,moriya2010}.  Further LC and spectral
modelling of stripped-envelope ECSNe are required to identify
transients corresponding to stripped-envelope ECSNe.

We are not the first to connect SN~2008ha and
ECSNe. \citet{pumo2009} made this link in the context of single-star
evolution model. We have shown here that stripped-envelope ECSNe can
also naturally arise from binary evolution.

Finally it is also tempting to compare these LCs to the
rapidly evolving transients found by \citet{drout2014}. While
the timescale of the transients and our
predicted rate of a few per cent is similar to those of these objects,
the luminosities here (around $-13~\mathrm{mag} -  -14~\mathrm{mag}$) are fainter than
those found in this population ($\sim -16~\mathrm{mag}$ or brighter).

\section{Conclusions}\label{sec:conclusions}
We have shown that stripped-envelope ECSNe can be observed as
rapidly-evolving faint transients. ECSNe are mostly considered to be
explosions of super-AGB stars, but stripped-envelope ECSNe can occur
especially from binary systems. The binary population synthesis model
from the BPASS code predict that ECSN fractions in core-collapse SNe
are 1.5 per cent ($Z=0.020$) and 8.6 per cent ($Z=0.004$). Among
ECSNe, 51 per cent ($Z=0.020$) and 14 per cent ($Z=0.004$) are predicted
to be stripped-envelope ECSNe
(Table~\ref{table:ecsnprop}). Stripped-envelope ECSNe typically have
ejecta masses of 0.3~\Msun\ ($Z=0.020$) and 0.6~\Msun\ ($Z=0.004$)
(Figure~\ref{fig:massdistribution}).

Assuming predicted explosion properties of ECSNe ($\Eej=1.5\times
10^{50}~\mathrm{erg}$ and $\Mni=2.5\times 10^{-3}~\Msun$), we find
that stripped-envelope ECSNe typically have the rise time of
around 7~days ($Z=0.020$) or 13~days ($Z=0.004$)
and peak luminosity of around $10^{41}~\mathrm{erg~s^{-1}}$ ($-13.8$~mag, $Z=0.020$)
or $7\times 10^{40}~\mathrm{erg~s^{-1}}$ ($-13.4$~mag, $Z=0.004$).
The LC properties are summarized in Figures~\ref{fig:lightcurves} and
\ref{fig:fractions}. Assuming the typical ejecta masses, we estimate
typical ejecta velocities of 7000~\kmps\ ($Z=0.020$) and 5000~\kmps\ ($Z=0.004$).

Expected LC properties and ejecta velocities are consistent with those
of SN~2008ha-like SNe. The stripped-envelope ECSN model can explain
the preference of SN~2008ha-like SNe to occur in low metallicity
environments. The possible red source detected at the location of
SN~2008ha could be a companion star of a stripped-envelope ECSN
progenitor. There has not been spectral modelling of stripped-envelope
ECSNe and it is required to identify them and determine if they match
SN~2008ha-like SNe or other transients.

\section*{Acknowledgements}
We thank the referee for the comments improved this paper.
This work is initiated during \textit{Electron Capture Supernovae \& Super-AGB
Star Workshop} held at Monash University, Australia.
TJM thanks Joe Lyman for comments.
TJM is supported by Japan Society for the Promotion of Science Postdoctoral Fellowships
for Research Abroad (26\textperiodcentered 51). JJE acknowledges
support from the University of Auckland. We recognize the vital
contribution of NeSI high-performance computing and the staff at the
Centre for eResearch at the University of Auckland. New Zealand's
national facilities are provided by the New Zealand eScience
Infrastructure (NeSI) and funded jointly by NeSI's collaborator
institutions and the Ministry of Business, Innovation and Employments
Infrastructure programme.










\bsp	
\label{lastpage}
\end{document}